\newcommand{\I}{Isabelle}
\newcommand{\myTitle}{Representation Theorems Obtained by Mining across Web Sources for Hints}
\newcommand{\restr}[2]{{\left. #1 \right|}_{#2}}
\newcommand{\im}[1]{#1^{*}}
\newcommand{\uim}[1]{#1^{\cup}}
\newcommand{\N}{\mathbb{N}}
\newcommand{\sdiff}{\backslash}
\newcommand{\emp}{\emptyset}
\newcommand{\pow}[1]{2^{#1}}
\newcommand{\fpow}[1]{{\overline{2}}^{#1}}
\newcommand{\biinv}[1]{\iota_{#1}}
\newcommand{\relcomp}[2]{#1 ; #2}
\newcommand{\Out}[3]{#1 - \left( #2, #3 \right)}
\newcommand{\out}[2]{#1 - #2}
\newcommand{\pointU}[2]{#1 + #2}
\newcommand{\Iff}{\leftrightarrow}
\newcommand{\olap}{\between}
\newcommand{\myOeis}{A284276}
\documentclass[conference]{IEEEtran}
\IEEEoverridecommandlockouts
\usepackage{cancel}
\usepackage{cite}
\usepackage{amsmath,amssymb,amsfonts,amsthm}
\usepackage{algorithmic}
\usepackage[inline]{enumitem}
\usepackage{graphicx}
\usepackage[]{hyperref}
\usepackage[]{microtype}
\usepackage{textcomp}
\usepackage{xcolor}
\def\BibTeX{{\rm B\kern-.05em{\sc i\kern-.025em b}\kern-.08em
    T\kern-.1667em\lower.7ex\hbox{E}\kern-.125emX}}

\DeclareMathOperator{\dom}{dom}
\DeclareMathOperator{\ran}{ran}
\DeclareMathOperator{\fie}{fie}
\DeclareMathOperator{\card}{card}
\DeclareMathOperator{\fx}{fx}
\DeclareMathOperator{\id}{\mathcal{I}}
\newtheorem{Cor}{Corollary}[section]
\newtheorem{Lm}[Cor]{Lemma}

\newtheorem{Thm}[Cor]{Theorem}
\newtheorem{Def}[Cor]{Definition}

\begin{document}

\title{\myTitle{}\\
\thanks{JKFB partially supported by the Austrian Science Fund under FWF Meitner project M-3338.
© 2022 IEEE.  Personal use of this material is permitted.  Permission from IEEE must be obtained for all other uses, in any current or future media, including reprinting/republishing this material for advertising or promotional purposes, creating new collective works, for resale or redistribution to servers or lists, or reuse of any copyrighted component of this work in other works.
}
}

\author{\IEEEauthorblockN{Marco B. Caminati}
\IEEEauthorblockA{\textit{School of Computing and Communications} \\
\textit{Lancaster University in Leipzig}\\
Leipzig, Germany\\
\url{orcid.org/0000-0002-4529-5442}}
\and
\IEEEauthorblockN{Juliana K. F. Bowles}
\IEEEauthorblockA{\textit{School of Computer Science, University of St Andrews}\\
KY16 9SX St Andrews, UK 
}
\IEEEauthorblockA{\textit{Software Competence Centre Hagenberg}\\ Softwarepark 32a, 4232 Hagenberg, Austria
\\ \url{orcid.org/0000-0002-5918-9114}
}
}

\maketitle

\begin{abstract}
A \emph{representation theorem} relates different mathematical structures by providing an isomorphism between them: that is, a one-to-one correspondence preserving their original properties.
Establishing that the two structures substantially behave in the same way, representation theorems typically provide insight and generate powerful techniques to study the involved structures, by cross-fertilising between the methodologies existing for each of the respective branches of mathematics.
When the related structures have no obvious a priori connection, however, such results can be, by their own nature, elusive.
Here, we show how data-mining across distinct web sources (including the Online Encyclopedia of Integer Sequences, OEIS), was crucial in the discovery of two original representation theorems relating \emph{event structures} (mathematical structures commonly used to represent concurrent discrete systems) to families of sets (endowed with elementary disjointness and subset relations) and to full graphs, respectively. 
The latter originally emerged in the apparently unrelated field of bioinformatics. 
As expected, our representation theorems are powerful, allowing to capitalise on existing theorems about full graphs to immediately conclude new facts about event structures.
Our contribution is twofold: on one hand, we illustrate our novel method to mine the web, resulting in thousands of candidate connections between distinct mathematical realms; on the other hand, we explore one of these connections to obtain our new representation theorems. 
We hope this paper can encourage people with relevant expertise to scrutinize these candidate connections.
We anticipate that, building on the ideas presented here, further connections can be unearthed, by refining the mining techniques and by extending the mined repositories.
\end{abstract}


\section{Introduction}
\label{RefSectIntro}
In automated mathematical discovery and experimental mathematics, a machine can be involved in any of the stages leading to the formulation of new mathematical conjectures
.
Usually, the interestingness and correctness of such conjectures are important criteria in informing how the machine performs its tasks.
Within this quite general framework, there is considerable variability as to the machine's role: it can, for example, generate conjectures~\cite{zeilberger1990holonomic,mccune1992automated,fajtlowicz1988conjectures}, attach to them a measure of interestingness~\cite{lenat1977automated}, search given input for plausible hints of conjectures~\cite{colton1999refactorable}, or compute results suggesting patterns that can inspire a mathematician~\cite{bailey2007experimental,davies2021advancing}.
Correspondingly, the degree of the machine's awareness of the involved mathematical objects varies from it applying a formal reasoning system on such objects to it merely examining examples of (possibly yet to be stated) conjectures.
We will focus on the latter end of this spectrum, sitting at the intersection between automated mathematical discovery and data mining. 
One obvious advantage of this choice is the extensive amount of data it grants: any conjecture involving finite objects (for example, graphs) leaves a trace obtained by counting the size of instances (for example, the number of vertices in graphs satisfying the hypotheses of a conjecture) of these objects.
These counts have a universal representation as decimal integers written in plain text, and 
therefore interesting matches between such counts can potentially be found over the vast range of all digitised documents. 
As a consequence, another advantage of this approach is that it is domain-agnostic 
and potentially able to link finite mathematical objects not apparently related (as long as one can count them), which we will see to be crucial in obtaining the results in this paper.

The idea is, therefore, to mine existing integer datasets for interesting relationships between them, possibly signaling deeper connections. 
This idea is by no means new~\cite{colton1999refactorable,colton2001mathematics,nguyenmining}. 
However, we believe that this paper will provide evidence that some aspects of it are worth more attention: the possibility of mining across distinct datasets and of exploiting datasets and tools
less specific to mathematics.

Section~\ref{RefSectMining} details how we put the above mining ideas into practice, and the resulting outcomes.
In the rest of the paper, we focus on one of these outcomes in particular, on the original mathematical results it hinted us to formulate, and on their proofs.
Section~\ref{RefSectBirkhoff} gives more specific, yet informal context about the family of theorems these results belong to and about their importance and methodological usefulness.
Section~\ref{RefSectNotations} introduces the definitions and notations to express these results.
Section~\ref{RefSectReprThm} illustrates a representation theorem for event structures (a computational model for discrete concurrent systems), Section~\ref{RefSectFull} introduces a theorem linking event structures to full graphs, and explains how both this result and that of Section~\ref{RefSectReprThm} were crucially suggested by the findings from Section~\ref{RefSectMining}.
Section~\ref{RefSectConcl} concludes.

\section{Mining Integer Sequences across Sources}
\label{RefSectMining}
The Online Encyclopedia of Integer Sequences (\href{http://oeis.org/}{OEIS})~\cite{sloane2013line} is a searchable online database containing the first terms (at least $4$, in decimal representation) of over $340,000$ integer sequences. Together with the field containing the terms, there are several meta-data fields: an unique ID, name, comments, references, keywords or flags (marking, for example, whether a sequence is finite), etc.
The OEIS has already been profitably used for research in automated mathematical discovery~\cite{ragni2011predicting,colton1999refactorable,holmstrom2017zeta}.
However, all the efforts we are aware of limit their discovery domain to the OEIS alone, potentially missing integer sequences not featured there.
This observation naturally leads one to investigate what can be found by looking up OEIS sequences (or fragments thereof) on the largest available repository of scientific literature, and Google Search (or Google, for short) is an obvious candidate: it indexes a huge number of web pages and documents and it subsumes Google Scholar, hosting an especially relevant subset of documents (i.e., scientific papers).
For our purposes, one particular attractiveness of Google Scholar is its own text extraction program~\cite{quint2008changes}, making analog scans of older papers searchable: papers older than OEIS are particularly at risk of having being omitted from it, and therefore worth being explored.

In 2019, Google Scholar was estimated, with $389$ million records, to be the largest bibliographic database~\cite{gusenbauer2019google}; 
by querying Google, we will have access to those records and many more. 
The price to pay for such a breadth of information is the inconvenience in accessing and processing it: while searching within the OEIS, one can automate numerical transformations on the sequences in order to facilitate matching between them.
This can happen either on the server side (typically through the Superseeker service~\cite{sloane2013line}) or on the user side~\cite{ragni2011predicting,colton1999refactorable,holmstrom2017zeta,nguyenmining}.
Under our approach, this possibility is largely gone, because any numerical transformation should happen before querying Google, leading to a multiplication of queries for every transformation: this is clearly impractical.
The transformations applied by Google on the queried terms are largely non-numerical (e.g., expanding a word into its English synonyms, or correcting possible mis-spellings) and hence immaterial in our case, except for possible formatting issues (e.g., matching the numerical representations $16000$ and $16,000$).
Furthermore, a bulk of noise results is to be expected, deriving from irrelevant occurrences of the searched numbers (e.g., in serial numbers, catalogs, etc.).

For these reasons, and since we have no control on how Google processes the input information it is passed, we need to carefully craft the format of that information beforehand. 
The main guiding idea in this task is simple: we want interesting matches between OEIS and Google, and complexity is a convenient measure of interestingness~\cite{colton2001mathematics}. 
Since the decimal representation length of an integer has a good correlation with its complexity (assuming non-significant figures are omitted, which is the case for the OEIS), we should ideally pass long integers from the OEIS to Google. 
This is especially true in our case where we need to treat, due to the limitations explained above, numbers as plain text, hence we do not have much else than length alone on which to base our assessment of the complexity (and therefore, of the interestingness) of a number.
However, we do not want too long numbers, because these are usually hard to compute, thereby potentially reducing too much the range of documents Google will return.
Therefore, we need to strike a balance with respect to the length of the numbers we pass to Google: we would like the minimal length leading to the exclusion of non-mathematical occurrences (such as dates, page numbers, catalog numbers, etc.) among the search results from Google. Empirically, we found that six digits do a reasonable job in that respect.

We downloaded all OEIS entries into a 16Gb SQLite database using~\cite{oeisTools},
removed all the sequences not having the ``hard'' keyword (meaning the sequence is not considered hard to compute), or having the field ``formula'' non empty (meaning that some mathematical property of the sequence is already known), or having no entries with more than five digits.
From the remaining entries, we sorted the terms according to their length, picked the smallest term with at least six digits and either the next one or (if there was no next one) the previous one. 
This scheme allowed us to produce, for 
$4123$ sequences, two distinct terms which were passed to Google, together with the directive 
\verb|-site:oeis.org|, to exclude matches within the OEIS.

The text snippets generated by Google in response, and describing the first matches among the documents indexed by it, were parsed as follows: first, the sequences with no matches were discarded, which left us with $3591$ sequences, all potentially interesting.
At this point, given the high number of matches to be manually examined, we decided to give priority for consideration to some matches, as follows.
We grepped each result for a set of arbitrary mathematical terms (including for example the words ``graph'', ``group'', ``ring''). 
If there was a match not occurring in the sequence OEIS name, that sequence was given priority.
Among those, the authors started from the ones pertaining fields where they felt most knowledgeable, and soon found an interesting pair: $41099$, $3528258 $, occurring both in OEIS \myOeis{} and in \cite[Section~4]{cowen1996enumeration}.
This match was decisive in suggesting the results we illustrate in the rest of this paper: it is an instance of Corollary~\ref{RefLmCard}, which, in turn, suggested us Theorems~\ref{RefLmBij} and~\ref{RefLmRepr} as dependencies.
Without that numeric cue, none of these results would have materialised: the theorems arose to explain why this match was not a coincidence.
The remaining matches need further human examination.

\section{Representation Theorems}
\label{RefSectBirkhoff}
A fundamental and extremely fruitful pattern in mathematics is to observe how some operations and correspondences between objects behave, and then to capture this behaviour via axioms, obtaining an abstract structure.
Together with the original meaning of the operations and correspondences one has thereby an abstract level: the structure axioms are formulas describing the formal relationship between objects, operations and correspondences, and can be manipulated, studied, and generalised algebraically without caring what their original meaning was. 
One can hence talk of two levels of thinking of the given mathematical objects: the original one (also called the concrete level), and the abstract one.

Examples of this way of obtaining abstract structures from concrete interpretations abound in mathematics: just to provide two well-known instances, from studying how permutations behave one obtains the group axioms; and from studying how $\cup$ and $\cap$ behave one obtains the (distributive) lattice axioms.

A natural question is how and to what extent one can go back from the abstract level to the concrete level: in other words, can any abstract structure be represented via a suitable concrete implementation of it?

For many important structures, this question is answered positively by \emph{representation theorems}, providing the existence of a suitable isomorphism allowing to go back and forth between these two levels;%
\footnote{In a more general acceptation, a representation theorem provides an isomorphism between an abstract structure and another structure, possibly itself abstract.} %
returning to the examples above, Cayley's representation theorem provides a representation of any group in terms of a permutation group~\cite[Section~II.7]{lane1999algebra}, and Birkhoff's representation theorem provides a representation of any finite distributive lattice in terms of a lattice of down-sets~\cite[Theorem~5.12]{davey2002introduction}. 

The fruitfulness of this two-level approach has many facets, including the ability of algebraically manipulating the concrete objects forgetting about their nature, thus seeing to what extent their known properties or relations are generalisable; or, oppositely, the reasoning aid given by a concrete setting as an inspiration to explore further consequences or generalisations of the abstract axioms given by properties of the concrete objects obeying them.
This fruitfulness is testified by the existence of dedicated fields using representations to study properties of given structures: e.g., representation theory studies the properties of groups using their representations as linear transformations of vector spaces.

In typical cases (such as the two just mentioned), the fact that the abstract level originated right from the start from the study of the concrete level makes such theorems quite natural to express and to prove: such results are, in these typical cases, attractively simple and elegant.%
\footnote{One should note that this simplicity is a boon with respect to the fruitfulness just mentioned.}

However, other mathematical structures could well have a more tortuous birth.
For example, prime event structures (formally introduced in Section~\ref{RefSectNotations}) historically and conceptually developed in stages: elementary event structures were expanded into prime event structures to accommodate nondeterminism~\cite[Section~2]{winskel1989introduction}.

As we will see in this paper, this tortuous birth led to miss, up to now, a remarkably simple representation theorem (\ref{RefLmRepr}) for prime event structures; whose simplicity, however, does not restrain the typical fertility of representation theorems, allowing us to immediately unearth unforeseen links between prime event structures and full graphs (the further representation theorem \ref{RefLmBij}) and cross fertilisation results (Corollaries \ref{RefLmCard} and \ref{RefLmCross}).
Another possible reason for this accident could be that the original purpose of prime event structures is to model computations of undetermined duration, which led to put less attention into the finite case, where our theorems are particularly simple; as briefly argued in Section~\ref{RefSectConcl}, we believe that Theorem~\ref{RefLmRepr}, besides its own importance, can serve as a fundamental stepping stone towards a generalisation to the infinite case. 
To give a final reason: given the original role of the elements in prime event structures as representatives of computational events, it is not natural to associate to them sets (as Theorem~\ref{RefLmRepr} does); or, at least, it is less natural than in cases, such as lattices, where a concrete level consisting of sets was historically a starting point to formulate the abstract level definitions.
The oversight of Theorem~\ref{RefLmRepr} is made even more surprising by the fact that other, more complicated representation theorems were formulated for prime event structures since their inception~\cite[Theorems~2.10 and 3.8]{winskel1989introduction}.

\section{Preliminaries and Event Structures}
\label{RefSectNotations}
Set membership, inclusion, union, intersection, set-theoretical difference, cartesian product
are denoted by the infix symbols $\in$, $\subseteq$, $\cup$, $\cap$, $\sdiff$, $\times$, respectively; 
arbitrary union and intersection over a set of sets are denoted by the prefix symbols $\bigcup$ and $\bigcap$. 
A set $R$ satisfying $R \subseteq X \times Y$ for some $X$, $Y$ (i.e., any set $R$ containing only ordered pairs) is called a binary relation or simply a relation. 
The minimal $X$ and $Y$ satisfying the previous inclusion are the domain ($\dom$) and range ($\ran$) of $R$, respectively, 
while its converse $R^{-1}$ is the set obtained by flipping the elements of each the pairs in $R$; the field of $R$ is $\fie R := \dom R \cup \ran R$. 
Given a set $X$, the restriction of $R$ to $X$ is defined as $\restr{R}{X}:=
\left( X \times \ran R \right) \cap R$, while the image of $X$ through $R$ is $\im R \left( X  \right) := \ran {\restr{R}{X}}$. 
The product or composition of relations $R$ and $S$ is defined as 
$
\relcomp R S := \left\{ \left( x, z \right).\ \exists y.\ \left( x,y \right) \in R \wedge \left( y, z \right) \in S \right\}.
$
$R$ is right-unique if, for any $x$, $\im R \left(  \left\{ x \right\}  \right)$ contains at most one element, while it is left-unique if $R^{-1}$ is right-unique. 
A right-unique relation is more commonly called a function or a map. 
In this case, there are special notations in use: 
\begin{enumerate*}
\item
$R \left( x \right)$, or even only $R\ x$, indicates the unique element of $\im R \left(  \left\{ x \right\}  \right)$, when $x \in \dom R$;
\item
$R: X \to Y$ indicates that $\dom R=X$ and that $\ran R \subseteq Y$;
\item
$X \ni x \overset{R}{\mapsto} y$ in lieu of $R=\left\{ \left( x,y \right). x \in X \right\}$, with ''$X \ni$`` or the superscript in $\overset{R}{\mapsto}$
\mbox{} possibly dropped when the context permits;
\item
$S \circ R$ in lieu of $\relcomp R S$.
\end{enumerate*}
A first example of a function is $\card$, associating to each set $X$ of a given family its unique cardinality $\card X$ (also denoted $\left| X \right|$). 
A left-unique function is called injective, or an injection.
$\pow X$ is the set of all subsets of $X$, while $\fpow{X} := \pow X \cap \im {\left( \card^{-1} \right)} \left( \N \right)$ denotes the finite subsets of $X$. 
Note that, for any relation $R$, $\im R$ is always a function
. 
When all the elements of $\ran R$ are sets,%
\footnote{This is always the case in some foundations: e.g., ZF, in which anything is a set.}
 there is an additional function one can derive from $R$: 
$\uim R := \dom R \ni x \mapsto \bigcup \im R \left(  \left\{ x \right\} \right) \subseteq \bigcup \ran R$, associating to each $x$ the union of all the sets in relation with $x$; if, furthermore, $R$ is a function, then $R$ and $\uim R$ coincide.
$\fx R := \fie \left( R \cap \id \right)=\dom \left( R \cap \id \right) = \ran \left( R \cap \id \right)$ is the set of fixed points of $R$, where $\id$ is the identity function.
A relation $R$ is said to be:
\begin{enumerate*}
\item
\emph{reflexive} if $\fie R \subseteq \fx R$;
\item
\emph{irreflexive} if $R \cap \id= \emp $;
\item
\emph{transitive} if $ \relcomp R R \subseteq R$;
\item
\emph{symmetric} if $R^{-1} \subseteq R$;
\item
\emph{antisymmetric} if $R \cap R^{-1} \subseteq \id$;
\item
a \emph{preorder} if it is both reflexive and transitive;
\item
a \emph{partial order} if it is an antisymmetric preorder.
\end{enumerate*}
A bijection between sets $X$ and $Y$ is an injection $f$ with $\dom f=X$ and $\ran f=Y$.

A prime event structure (or just \emph{event structure}, ES)~\cite{winskel1989introduction} models a concurrent computation by specifying which computational events are causally dependent and which events mutually exclude.
This is attained by two relations $\le$ (causality), and $\#$ (conflict) as from the following definition.
\begin{Def}
\label{RefDefEs}
An event structure is a pair of relations $ \left( \leq, \# \right)$ where $\leq$ is a partial order, $\#$ is irreflexive and symmetric, 
$ \left( \fie \leq  \right) \supseteq \left(  \fie \#  \right)$ 
is called the set of events, and for any three events $x_0, x_1, y$: $ x_0 \# y \wedge x_0 \leq x_1 \to x_1 \# y$. 
\end{Def}
The last condition is referred to as conflict propagation.
The standard infix notation in Definition~\ref{RefDefEs} can get cumbersome, therefore we will often use the set theoretical notation and denote these relations with letters, for example writing $\left( x,y \right) \in D$ in lieu of $x \leq y$ and $\left( x,y \right) \in U$ in lieu of $x \# y$.

\begin{figure}[htbp]
\centering{}
\includegraphics[scale=.8]{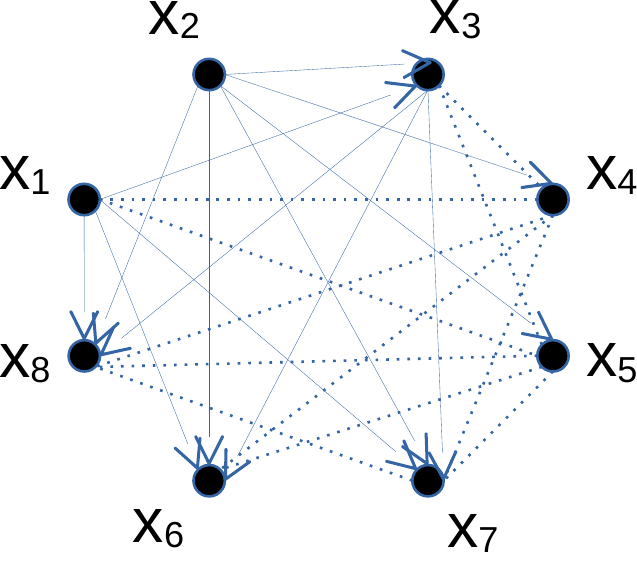}
\caption{An example event structure, with eight events related by causality (denoted by an arrow standing for $\leq$) and conflict (denoted by a dashed line).}
\label{RefFigEs}
\end{figure}

\section{A Representation Theorem for ESs}
\label{RefSectReprThm}
The main result of this section is Theorem~\ref{RefLmRepr}, establishing that elements of any finite ES can be represented as finite sets, in such a way that $\leq$ corresponds to $\supseteq$ and $\#$ to disjointness.
Formally, this means that it is always possible to find a function $f$ associating a set to each event of a finite ES subject to the constraints given by Definition~\ref{RefDefRepr}. 
We will call such a function a \emph{representation} for the given ES.

\begin{Def}
\label{RefDefRepr}
Given two binary relations $D$ and $U$, the set-valued function $f$ is a \emph{representation} for $(D,U)$ if
\begin{align}
\label{RefFmReprSub}
\forall x \ y \in \dom f.\ &\left( \left( x, y \right) \in D \leftrightarrow f \left( x \right) \supseteq f \left( y \right) \right) 
\ \wedge \\ 
\forall x \ y \in \dom f.\ 
& \label{RefFmReprDisj}
\left( 
\left( x, y \right) \in U \leftrightarrow f \left( x \right) \cap f \left( y \right) = \emp
\right)
. &
\end{align}
\end{Def}

We are now ready to state our representation theorem.

\begin{Thm}[Representation theorem]
\label{RefLmRepr}
Consider two binary relations $D$ and $U$, with $D$ finite and $\fie U \subseteq \fie D$. Then $\left( D, U \right)$ is an event structure if and only if there is an injective representation $f : \fie D \to \fpow{\N}\sdiff \left\{ \emp \right\}$ for $(D, U)$.
\end{Thm}

That is, a sufficient and necessary condition for a given finite number of events to form an event structure is the possibility of associating to each of them a set in such a way that $\supseteq$ corresponds to $\to^*$ and $\#$ corresponds to disjointness. 
In the theorem, the associated sets are all subsets of $\N$; however, any other infinite superset would do: the choice of $\N$ is only dictated by technical convenience. 

\begin{figure}[h]
\centering{}
\includegraphics[scale=.8]{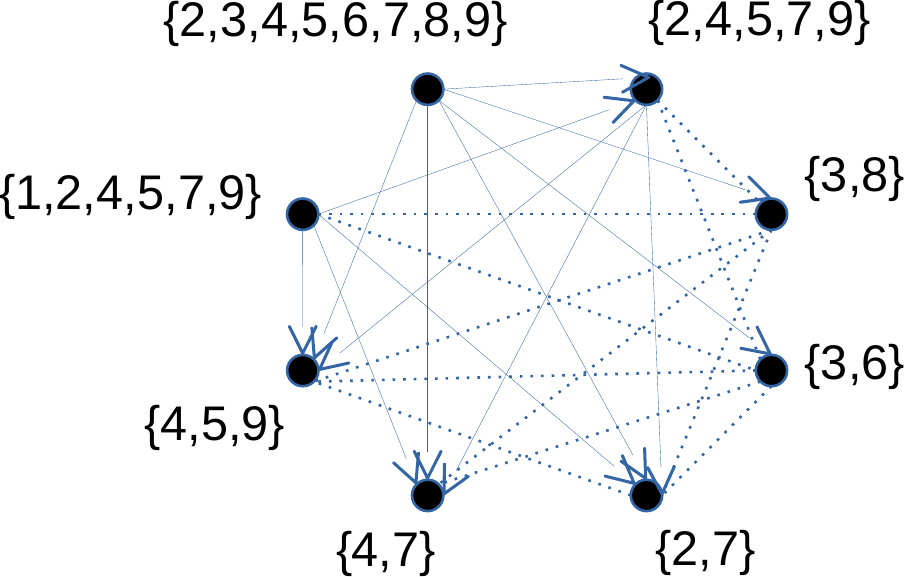}
\caption{A representation for the event structure of Figure~\ref{RefFigEs}. Now, the arrows represent $\supseteq$ and the dashed lines the disjointness relation. Theorem~\ref{RefLmRepr} states that any set of events is an event structure if and only if such a representation is constructible.}.
\label{RefFigRepr}
\end{figure}

Figure~\ref{RefFigRepr} shows a representation for the ES of Figure~\ref{RefFigEs}.

The two implications composing the logical equivalence (``if and only if'') in Theorem~\ref{RefLmRepr} are proved separately in Sections~\ref{RefSectReprIEs} and~\ref{RefSectEsIRepr}.

\subsection{Having a Representation Implies Being an ES}
\label{RefSectReprIEs}
The first step is proving that condition~\eqref{RefFmReprSub} is strong enough to impose the partial order properties of $\supseteq$ onto $D$. 
This can be done directly but, instead, we will break down the proof into more general results, which we will gather in Lemma~\ref{RefLmPreserving}.
Formula~\eqref{RefFmReprSub} closely resembles the definition of $f$ being an order embedding~\cite
{davey2002introduction}, except for the fact that here $D$ is not assumed to be a partial order (because this is what we need to prove), while the standard definition of an order embedding takes that as a pre-condition.
Therefore, we take the chance to study what can be proven about two relations linked by an order embedding when we drop basic assumptions.
In this section, we reason about generic relations $P$ and $Q$, rather than the specific ones, $D$ and $\supseteq$, appearing in~\eqref{RefFmReprSub}. 
We start by stating the standard definitions of order-preserving and order-embedding, only with the order assumptions dropped, together with some additional definitions.
\begin{Def}
\label{RefDefPreserve}
Given two relations $P$ and $Q$, a map $f$ is said to be 
\begin{enumerate*}
\item \label{RefDefPreservePreserve}
$\left( P, Q \right)$-preserving if $ \forall x_0, \ x_1 \in \dom f. \left( x_0, x_1 \right) \in P \to 
\left( f \left( x_0 \right), f \left( x_1 \right) \right) \in Q$;
\item \label{RefDefPreserveConverse}
$\left( P, Q \right)$-converse-preserving if $ \forall x_0, \ x_1 \in \dom f. $ 
$\left( f \left( x_0 \right), f \left( x_1 \right) \right) \in Q$ 
 $\to
\left( x_0, x_1 \right) \in P;
$
\item \label{RefDefPreserveEmbed}
a $\left( P, Q \right)$-embedding if it is both $\left( P, Q \right)$-preserving and $\left( P, Q \right)$-converse-preserving.
\end{enumerate*}
The prefix ``$\left( P, Q \right)$-'' can be dropped when no ambiguity arises.
We also introduce the map $\biinv f := \left( y_0, y_1 \right) \mapsto \im {\left(f^{-1}\right)} \left\{ y_0 \right\} \times \im {\left(f^{-1}\right)} \left\{ y_1 \right\}$.
\end{Def}
\begin{Lm}
\label{RefLmPreserving}
Let $P$, $Q$ be relations, $f$ a function.
\begin{enumerate*}
\item \label{RefLmPreserving47r}
$f$ is converse-preserving iff $\bigcup \im {\biinv f}\ Q \subseteq P$; 
\item
\label{RefLmPreserving48k}
$\im{\left( f^{-1} \right)} \left( \fx Q \right) \subseteq \fx \left( \bigcup \im{\biinv f} \ Q \right)$. 
\item
\label{RefLmPreserving47u}
$\fie P \subseteq \dom f \to$. 
 $\left(  f \text{ is preserving iff } P \subseteq \bigcup \im {\biinv f}\  Q \right)$. 
\item
\label{RefLmPreserving48l}
If $Q$ is transitive, then $\bigcup \im {\biinv f}\ Q$ is. 
\item
\label{RefLmPreserving48kk}
If $Q$ is reflexive, then $\fie \left( \bigcup \im{\biinv f} \ Q \right) \subseteq  
\im{\left( f^{-1} \right)} \left( \fx Q \right)$. 
\end{enumerate*}
\end{Lm}

\begin{proof}
Theses~\eqref{RefLmPreserving47r} and~\eqref{RefLmPreserving47u} are easy rephrasings of, respectively, \eqref{RefDefPreserveConverse} and~\eqref{RefDefPreservePreserve} in Definition~\ref{RefDefPreserve}.
Now set $P' := \bigcup \im {\biinv f}\ Q$.
Proof of~\eqref{RefLmPreserving48k}: if $\left( y_0, y_0 \right) \in Q$ and $x_0 \in \im {\left(f^{-1}\right)} \left\{ y_0 \right\}$, then, in particular, $ \left( x_0, x_0 \right) \in \im {\left(f^{-1}\right)} \left\{ y_0 \right\}
\times \im {\left(f^{-1}\right)} \left\{ y_0 \right\} \subseteq P'$.
Proof of~\eqref{RefLmPreserving48l}: 
consider $\left( x_0, x1 \right), \left( x_1, x_2 \right) \in P'$; 
$\left\{ (f\ x_0, f\ x_1), \left( f\ x_1, f\ x_2 \right)  \right\} \subseteq Q$, so that 
$\left( f\ x_0, f\ x_2 \right) \in Q$ by transitivity, and $\left( x_0, x_2 \right) \in
\im {\left(f^{-1}\right)} \left\{ f\ x_0 \right\}
\times
\im {\left(f^{-1}\right)} \left\{ f\ x_2 \right\} \subseteq P'.
$
Proof of~\eqref{RefLmPreserving48kk}: by construction of $P'$, $x_0 \in \fie P'$ implies the existence of $y_0 \in \fie Q$ such that 
$x_0 \in \im {\left(f^{-1}\right)} \left\{ y_0 \right\}$. Now, by reflexivity of $Q$:
$ P' \supseteq \im {\left(f^{-1}\right)} \left\{ y_0 \right\} 
\times
\im {\left(f^{-1}\right)} \left\{ y_0 \right\} \ni \left( x_0, x_0 \right).$
\end{proof}

\begin{Cor}
\label{RefLmPreorder}
Assume $f$ is a $\left( P, Q \right)$-embedding, $\fie P \subseteq \dom f$.
If $Q$ is a preorder, then $P$ is.
Moreover, if $f$ is injective and defined over $\fie P$, and $Q$ is a partial order, 
then $P$ is.
\end{Cor}
\begin{proof}
$P' := \bigcup \im {\biinv f}\ Q$ inherits $Q$'s transitivity by virtue of~\eqref{RefLmPreserving48l} in Lemma~\ref{RefLmPreserving}, and $Q$'s reflexivity by chaining~\eqref{RefLmPreserving48kk} and~\eqref{RefLmPreserving48k} of Lemma~\ref{RefLmPreserving}.
Using~\eqref{RefLmPreserving47r} and~\eqref{RefLmPreserving47u} in Lemma~\ref{RefLmPreserving}, the embedding property of $f$ implies $P=P'$, and we just saw that $P'$ is a preorder.
Assume $\left\{ \left( x, y \right), \left( y,x \right)  \right\} \subseteq P$.
Then $f\ x=f\ y$ by antisymmetry of $Q$, so that the antisymmetry of $P$ is satisfied by injectivity.
\end{proof}

\begin{Lm} 
\label{RefLmMain1}
Assume $f$ is an injective representation $f : \fie D \to \fpow{\N}\sdiff \left\{ \emp \right\}$ for $(D, U)$. Then $\left( D, U \right)$ is an ES.
\end{Lm}
\begin{proof}
\eqref{RefFmReprSub} means that $f$ is a $\left( D, \supseteq \right)$-embedding, and the latter is a partial order, so that $D$ also is by virtue of Corollary~\ref{RefLmPreorder}.
Consider events $x_0, x_1, y$, and assume $\left(  x_0, y \right) \in U \wedge \left( x_0, x_1 \right) \in D$.
Then $f\ x_0 \cap f\ y = \emp \wedge f\ x_0 \supseteq f\ x_1$, giving conflict propagation.
The symmetry of $U$ is immediate from that of $\cap$, and the irreflexivity of $U$ uses $\emp \notin \ran f$.
\end{proof}

\subsection{Any ES Has a Representation}
\label{RefSectEsIRepr}
The proof of this direction (the ``only if'' part of Theorem~\ref{RefLmRepr}) is more elaborate than 
the other one (Lemma~\ref{RefLmMain1}), because we now need to construct a representation $f$ given any finite event structure.
We will do that recursively: we will remove one suitable element of the given event structure, thus lowering its cardinality and obtaining a representation for this reduced event structure, and we will show how to extend this representation so as its property of being a representation still holds with respect to the original event structure.
The aforementioned operations of removing one element from a relation and of extension of a function are formally introduced, in forms suitable for our goals, in Definition~\ref{RefDefUnion}.

\begin{Def}
\label{RefDefUnion}
The subtraction of sets $X$, $Y$ from the relation $R$ is defined as
$\Out R X Y := R \sdiff ((X \times \ran R) \cup (\dom R \times Y))$.
We will use the shorthand notation $\out R s$ to indicate 
$\Out R {\left\{ s \right\}} {\left\{ s \right\}}$.
The pointwise union of relations $R_0$ and $R_1$ is the function 
$\pointU R_0 R_1 := 
\uim {\left( R_0 \cup R_1 \right)}
$.
By associativity, one extends this notion to multiple relations in the obvious way, writing
$ \sum_i R_i $.
For singleton relations, we can write, e.g., 
$\pointU R {\left( x,y \right)} $ in lieu of $\pointU R {\left\{ \left( x,y \right) \right\}}$.
\end{Def}

The following lemma gives conditions under which we can extend a representation into one having a larger domain.

\begin{Lm} 
\label{RefLmExtend}
Let $g$ be a representation for $\left( \out D s, \out U s \right) $.
Assume that $\im D { \left\{ s \right\}} = \left\{ s \right\} \not \subseteq \im U \left\{ s \right\} \cup \dom g$, and that
$\forall x \in \dom g. \ \left( x, s \right) \in U \Iff \left( s,x \right) \in U.$
If, for any $x \in \dom g$, the non empty set $Y$ satisfies all the following properties:
\begin{enumerate*}
\item
\label{RefFormulaFresh}
$g\ x \not \subseteq Y$,
\item
\label{RefFormulaRequirementSupset}
$Y \subseteq g\ x \Iff x \in \im {\left( D^{-1}  \right)} \left\{ s \right\} \sdiff \left\{ s \right\}$,
\item
\label{RefFormulaRequirementDisjoint}
$
g\ x \cap Y = \emp \Iff x \in \im {\left( U^{-1} \right)} \left\{ s \right\},
$
\end{enumerate*}
then $\pointU g \left( s, Y \right)$ is a representation for $ \left( D, U \right)$.
\end{Lm}

\begin{proof}
$f:= g + \left( s, Y \right)$ extends $g$, therefore we only need to check conditions~\eqref{RefFmReprSub} and~\eqref{RefFmReprDisj} of Definition~\ref{RefDefRepr} in the case $s \in \left\{ x, y \right\}$. 
What is more, the first of these conditions is trivial when $x=s$, so that we only need to check the case $y=s, x \neq s$, which immediately gives, using hypothesis~\ref{RefFormulaRequirementSupset}:
$ \left( x,s \right) \in D \leftrightarrow x \in \im{\left( D^{-1} \right)} \ \left\{ s \right\} \sdiff \left\{ s \right\} 
\leftrightarrow f\ s = Y \subseteq g\ x = f\ x.
$
To check formula~\eqref{RefFmReprDisj} of Definition~\ref{RefDefRepr} in the same case we use hypothesis~\ref{RefFormulaRequirementDisjoint}:
$
\left( x, s \right) \in U \leftrightarrow x \in
\im {\left( U^{-1} \right)}\ \left\{ s \right\} \leftrightarrow g\ x \cap Y = \emp \leftrightarrow 
f\ x \cap f\ s = \emp,
$
where the last step employed hypothesis~\ref{RefFormulaFresh}.
A symmetric argument concludes the proof by showing the same formula in the case $x=s, y \neq s$. 
\end{proof}

While condition~\eqref{RefFormulaFresh} in Lemma~\ref{RefLmExtend} merely requires that $Y$ is ``fresh'', and is therefore usually easy to meet, not every representation $f$ admits a $Y$ satisfying the remaining conditions~ \eqref{RefFormulaRequirementSupset} and~\eqref{RefFormulaRequirementDisjoint}. 
However, it is always possible to augment a representation $f$ to make this happen, where by ``augmenting'' we mean the action of enlarging the sets in $\ran f$.
This is detailed by Lemma~\ref{RefLmIntermediate}.

\begin{Lm} 
\label{RefLmIntermediate}
Consider two relations $D$, $U$,  with 
$
\left( 
\im {\left( D^{-1} \right)} \left\{ s \right\} \sdiff \left\{ s \right\} 
 \right)
\cap 
\im {\left( U^{-1} \right)} \left\{ s \right\} = \emp
$ for some fixed $s$, 
a set-valued function $f$, and a finite list
$g_i := X_i \times \left\{ y_i \right\} , i=1, \ldots, n$ of constant, non-empty functions. 
Let $g:= f + \sum g_i + \left( 
\left( \im {\left( D^{-1} \right)} \left\{ s \right\} \sdiff \left\{ s \right\} \right)
\times \left\{ y \right\}
\right)$ and 
$
Y := y \cup \bigcup_{i=1}^{n} y_i
,
$
where $y$ is a set not included in $\bigcup y_i \cup \bigcup \ran f$.
Assume
\begin{enumerate*}
\item
\label{RefLmIntermediate01}
$U^{-1} \left\{ s \right\} \cap \bigcup X_i = \emp;$
\item
\label{RefLmIntermediate02}
$ \im {\left( D^{-1} \right)}
\left\{ s \right\} \sdiff \left\{ s \right\} \subseteq \bigcap X_i ;$ 
\item
\label{RefLmIntermediate03}
$\dom g \sdiff \left( 
\im {\left( D^{-1} \right)} \left\{ s \right\}
\cup
\im {\left( U^{-1} \right)} \left\{ s \right\}
\right) \subseteq \bigcup_{i=1}^{n} X_i;$
\item
\label{RefLmIntermediate04}
$s \notin \dom g;$
\item
\label{RefLmIntermediate05}
$Y \cap \left( \left\{ \emp \right\} \cup \bigcup \ran f  \right)=\emp.$
\end{enumerate*}
Then one has, for any $x \in \dom g$:
$Y \subseteq g\ x \Iff x \in \im {\left( D^{-1}  \right)} \left\{ s \right\} \sdiff \left\{ s \right\}$, and
$
g\ x \cap Y = \emp \Iff x \in \im {\left( U^{-1} \right)} \left\{ s \right\}.
$
\end{Lm}

\begin{proof}
Let $S:=\im {\left( D^{-1}  \right)} \left\{ s \right\} \sdiff \left\{ s \right\}$ and fix $x \in \dom g$. Assume $ Y \subseteq g\ x$. Then, in particular, $y \subseteq g\ x$ and, since $y$ is fresh, 
it must be $x \in S$ by construction of $g$. Conversely, assume $x \in S$.
By hypothesis~\eqref{RefLmIntermediate02}, then, each $y_i$ must be included in $g\ x$, as is $y$, finishing the proof of the first thesis.
Now assume $g\ x \cap Y = \emp$. Then $ x \notin S \cup \bigcup X_i $ by construction of $g$ and $Y$, which yields 
$ x \in \im {\left( U^{-1} \right)} \left\{ s \right\}$
by \eqref{RefLmIntermediate03} and \eqref{RefLmIntermediate04}.
Finally, assume $x \in \im {\left( U^{-1} \right)} \left\{ s \right\}$.
From~\eqref{RefLmIntermediate01} and~\eqref{RefLmIntermediate02}, one draws $g\ x = f\ x$, completing the proof by virtue of~\eqref{RefLmIntermediate05}. 
\end{proof}

We note that the requirements on $D$, $U$, and $f$ in the last lemma are weaker than what we will need: for example, $f$ is not required to be a representation, or $D$ to be a partial order.
To obtain the final result in this section, we now just need to pipe Lemma~\ref{RefLmIntermediate} into Lemma~\ref{RefLmExtend}; to do so, we want to make sure that, referring to Lemma~\ref{RefLmIntermediate}, when $f$ is a representation, so $g$ is. 
The next result gives guidance in picking the $X_i$'s in Lemma~\ref{RefLmIntermediate} to attain this, after which we will be in a position to give the proof of Theorem~\ref{RefLmRepr}, including in particular the result that any finite ES has a representation.

\begin{Lm}
\label{RefLmAugment} 
Let $f$ be a representation for $\left( D, U \right)$, and $f'$ be a map with $\dom f' \subseteq \dom f$ and 
$ \left( \bigcup \ran f \right) \cap \bigcup \ran f' = \emp$.
Assume that, for any $x \in \dom f$:
\begin{enumerate*}
\item \label{RefLmAugmentH1}
$ \forall y \in \dom f. \ \left( x, y \right) \in U \to $ $\left(x \neq y \wedge \card \left( \dom f' \cap \left\{ x, y \right\} \leq 1 \right) \right)$ and
\item \label{RefLmAugmentH2}
$
\forall y \in \dom f'. \left( x, y \right) \in D \to $ $\left( x \in \dom f' \wedge f'\ x \supseteq f'\ y \right).
$
\end{enumerate*}
Then $f + f'$ is also a representation for $ \left( D, U \right)$.
\end{Lm}

\begin{proof}
Let $g:=f+f'$ and fix $x, y \in \dom g = \dom f$.
Assume $\left( x, y \right) \in D$ and $g\ x \not \supseteq g\ y$. Then, by construction of $g$ and by hypothesis~\ref{RefLmAugmentH2}, using the monotonicity of $\cup$, one concludes $y \in \dom f'$ and $x\in \dom f \sdiff \dom f'$, which contradicts hypothesis~\ref{RefLmAugmentH2}.
Viceversa, assume $g\ x \supseteq g\ y$; then $f\ x \supseteq f\ y$ using
$ \left( \bigcup \ran f \right) \cap \bigcup \ran f' = \emp$, so that $\left( x,y \right) \in D$ by representativity of $f$.
Now assume $\left( x,y \right) \in U$. Then $g\ x \cap g\ y = \left( f\ x \cup X \right) \cap \left( f\ y \cup Y \right)$ where at least one among $X$ and $Y$ is empty, thanks to hypothesis~\ref{RefLmAugmentH1}.
Since $X \cup Y \subseteq \bigcup \ran f'$, which is disjoint from $\bigcup \ran f \supseteq \left(  f\ x \cup f\ y  \right)$, one obtains $g\ x \cap g\ y=f\ x \cap f\ y= \emp$ by representativity of $f$.
Finally, assume $g\ x \cap g\ y = \emp$; in particular, $f\ x \cap f\ y= \emp$, yielding $\left( x, y \right) \in U$ again by representativity of $f$. 
\end{proof}

\begin{proof}[Proof of Theorem~\ref{RefLmRepr}]
One direction is given by Lemma~\ref{RefLmMain1}.
For the converse, assume the existence of finite event structures not admitting an injective representation 
$\fie D \to \fpow{\N}\sdiff \left\{ \emp \right\}$.
Among such counterexamples, we can take one (let us denote it $\left( D, U \right)$, with $\fie U \subseteq \fie D$) whose causality relation $D$ has minimal cardinality.
It is immediate to check that $D$ cannot be empty: one consequence of this is that we can fix a $D$-maximal element $s$ of it (due to $D$ being finite); another consequence is that $\card \left( \out D s \right) < \card\ D$.
Moreover, $\left( d:= \out D s, u:= \out U s \right)$ is still an event structure, and $\fie u \subseteq \fie d$, so that we can obtain a representation for it: $f: \fie d \to \fpow{\N}\sdiff \left\{ \emp \right\}$.
We now need to apply Lemma~\ref{RefLmAugment} to $f$, in order to obtain another representation over the same domain to which to apply Lemma~\ref{RefLmIntermediate}.
To this end, let us consider the set of events concurrent to $s$: 
$ C = \fie D - D^{-1} \left\{ s \right\} - U^{-1} \left\{ s \right\} $, together with a list of non-empty sets $\left\{ Z_i.\ i=1, \ldots, n \right\}$ being conflict-free and downward-closed, and covering $C$.
This is always possible, for example by taking $ \left\{ D^{-1} \left\{ c \right\}.\ c \in C \right\}$.
Finally, define $X_i := Z_i \cup D^{-1}\left\{ s \right\} \sdiff \left\{ s \right\}$.
Note that each $X_i$ is still conflict-free, which implies, together with the irreflexivity of $U$, hypothesis~\eqref{RefLmAugmentH1} of Lemma~\ref{RefLmAugment}.
Now we construct the constant functions $g_i := X_i \times \left\{ m+i \right\}$, where $m$ is any fixed natural $> \max \bigcup \ran f$. 
The fact that each $X_i$ is still, as is $Z_i$, downward-closed, together with the way we constructed $g_i$, allows each $g_i$ to satisfy hypothesis~\eqref{RefLmAugmentH2} of Lemma~\ref{RefLmAugment}.
Therefore, $f+g_1$ is also a representation for $\left( d, u \right)$ and, iterating this reasoning, so is
$f + \sum g_i$. 
The same reasoning can now be applied to $D^{-1} \left\{ s \right\} - \left\{ s \right\} \times \left\{ m+n+1 \right\}$, so that $g := f + \sum g_i + D^{-1} \left\{ s \right\} - \left\{ s \right\} \times \left\{ m+n+1 \right\}$ is still a representation for $\left( d, u \right)$.
Setting $Y := \left\{ m, \ldots, m+n+1 \right\}$, it is easy to check that $D$, $U$, $f$, $g$, $Y$ and the $X_i$'s satisfy all of Lemma~\ref{RefLmIntermediate}'s  hypotheses, so that
$Y \subseteq g\ x \Iff x \in \im {\left( D^{-1}  \right)} \left\{ s \right\} \sdiff \left\{ s \right\}$
and
$g\ x \cap Y = \emp \Iff x \in \im {\left( U^{-1} \right)} \left\{ s \right\}.$
Moreover, since $Y$ is fresh and $ \emp \notin \ran f$, we also have $g\ x \not \subseteq Y$ for every $x \in \dom g=\dom f$. 
Therefore, by Lemma~\ref{RefLmExtend}, $h := g \cup \left\{ \left( s, Y \right) \right\}$ is a representation for $\left( D, U \right)$.
Finally, it is straightforward to see, since $Y \cap \bigcup \ran f = \emp$, that $g$ inherits the injectivity of $f$ and, therefore, that $g \cup \left\{ \left( s, Y \right) \right\}$ is also injective.
It is also immediate to see that $\emp \notin \bigcup \ran h$, due to $Y \neq \emp$.
We thus reached a contradiction with our assumption that $\left( D, U \right)$ admitted no such injective representation.
\end{proof}

\section{Full Graphs}
\label{RefSectFull}
Given a family of sets, one can construct a graph where each vertex corresponds to a set, a directed edge links supersets to subsets, and an undirected one connects overlapping sets.
Such a construction arises when computationally facing the question of whether subelements of genes are linked together in a linear order~\cite{fulkerson1965incidence}.
Definition~\ref{RefDefFull} formally specify the graphs which can be built in this manner.
\begin{Def}
\label{RefDefFull}
A \emph{full graph} is a mixed, unweighted, simple graph over vertices $V$, of directed edges $D$, and undirected edges $T$ such that there is an injective function $f$ on $V$ yielding non-empty sets and with the property
$\forall x, \ y \in V.\ 
\left( 
\left( x, y \right) \in D \leftrightarrow f\ x \supseteq f\ y\  \right) \wedge$
$
\left( 
\left( x, y \right) \in T \leftrightarrow f\ x \text{ and } f\ y \text{ overlap} \right);$ 
here, we say that two sets $A$ and $B$ \emph{overlap} (written $ A \olap B $) when $A \cap B \notin \left\{ A, B, \emp \right\}$.
We call $f$ a \emph{fg-representation} of the full graph $(D,T)$.
Alternatively, we will say that $T$ makes a full graph of $D$ (through $f$) when such an fg-representation $f$ exists.
Similarly, we will say that a relation $U$ is admissible for $D$ (through $f$) when $\fie U \subseteq \fie D$ and there is a similar $f$ being a representation (as from Definition~\ref{RefDefRepr}) for $\left( D, U \right)$.
\end{Def}
Note that an undirected edge linking $x$ and $y$ is represented by two pairs $\left( x, y \right)$ and $\left( y,x \right)$ in $T$. While redundant, this representation allows us to formally consider $T$ a (symmetric) relation, so that any full graph can be adequately represented by a pair $\left( D, T \right)$ of relations, also thanks to the fact that it is simple (e.g., without multiple edges).
We can omit $V$ because any full graph must have a loop on every vertex, so that $V = \fie D$ is redundant.

Theorem~\ref{RefLmBij} is our second representation theorem for event structures, providing a bijective construction relating them to full graphs.

\begin{Thm}
\label{RefLmBij}
Consider a finite relation $D$ and $F_D:=R \mapsto \left( \fie D \times \fie D \right) \sdiff \left( D \cup D^{-1} \right) \sdiff R$. A bijection between 
$X:=\left\{ T | T \text{ makes a full graph of } D \right\} $ and 
$Y:=\left\{ U | U \text{ is admissible for } D \right\} $ is given by $\restr {F_D} {X}$.
\end{Thm}

\begin{figure}[htbp]
\centering{}
\includegraphics[scale=.8]{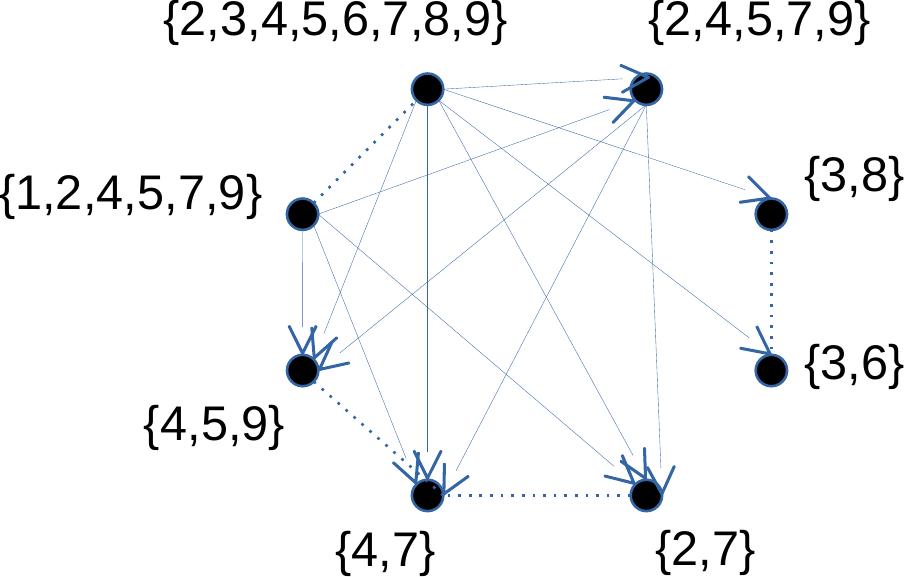}
\caption{By applying the inverse of $F_D$ appearing in Theorem~\ref{RefLmBij} to the event structure of Figures~\ref{RefFigEs} and \ref{RefFigRepr}, one obtains the full graph example originally featured in Section~3 of \cite{fulkerson1965incidence}. Here, the arrows represent $\supseteq$, and the dashed lines the overlapping relation.}
\label{RefFigFg}
\end{figure}

Figure~\ref{RefFigFg} shows the application of Theorem~\ref{RefLmBij} to the event structure of Figures~\ref{RefFigEs} and \ref{RefFigRepr}.

\begin{proof}
Writing just $F$ for $F_D$, it suffices to show four claims: $\restr F X$ is injective, $\restr F Y$ is injective, $\im F\ X \subseteq Y$ and $\im F\ Y \subseteq X$. 
The injectivity claims follow from the general fact that $ \restr 
{\left( v \mapsto w \sdiff v \right)} {\pow w} 
$ is always injective, and from $X \cup Y \subseteq \pow {
\left( \fie D \times \fie D \right) \sdiff \left( D \cup D^{-1} \right)
}$. 
In turn, the last inclusion follows from the fact that a relation admissible for $D$ is necessarily disjoint from $D \cup D^{-1}$ and similarly for one making a full graph of $D$. 
For the third claim: consider $T$ making a full graph of $D$ through $f$, and vertices $x,\ y$; now 
$ 
f\ x \cap f\ y = \emp \leftrightarrow 
f\ x \cancel \supseteq f\ y \wedge f\ y \cancel \supseteq f\ x \wedge f\ x \cancel \olap f\ y
\leftrightarrow 
\left( x, y \right) \notin D \cup D^{-1} \cup T 
\leftrightarrow \left( x, y \right) \in F\ T,
$ so that $F\ T$ is admissible for $D$ through the same $f$ (with the part $\fie \left( F\ T \right) \subseteq \fie D$ being straightforward).
Similarly for the last claim.
\end{proof}

\begin{Cor}
\label{RefLmCard}
Consider a finite set $V$, and the set $P$ of partial orders having field $V$.
The sets $E\left( V \right)$ and $F \left( V \right)$ of event structures over $V$ and of full graphs over $V$, respectively, are given by
$E \left( V \right) = \bigcup_{D \in P} \left\{ D \right\} \times \left\{ U| \ U \text{ is admissible for } D \right\},$
$F \left( V \right) = \bigcup_{D \in P} \left\{ D \right\} \times \left\{ T| \ T \text{ makes a full graph of } D \right\}.$
They have the same cardinality.
\end{Cor}

\begin{proof}
The first equality follows from Theorem~\ref{RefLmRepr}, while the second is a rephrasing of Definition~\ref{RefDefFull}.
Both feature disjoint unions, so that the cardinality claim follows from~\ref{RefLmBij}.
\end{proof}

Corollary~\ref{RefLmCard} reveals why the match between OEIS~\myOeis{} and the countings in the paper~\cite{cowen1996enumeration}, found by querying Google with OEIS minings, is not a coincidence: the former counts event structures over sets of given cardinalities, and the latter does the same for full graphs.
This correspondence between two previously detached world immediately yields new results by translating existing theorems previously applied to only one world. The next corollary lists only two, among the easiest, of them.

\begin{Cor}
\label{RefLmCross}
\begin{itemize*}
\item
Exactly $561658287$ full graphs are constructable on seven vertices, including isomorphic ones.
\item
$ \lim_{\left| V \right|  \to \infty} \frac{\log_2 \left|  E \left( V \right)  \right| }{{\left| V \right|}^2}=\frac{1}{2},$
\end{itemize*}
where $E$ is defined as in~\ref{RefLmCard}. 
\end{Cor}

\begin{proof}
Immediate by applying~\ref{RefLmCard} to OEIS~\myOeis{} and to the main result of~\cite{kleitman1995asymptotic}, respectively.
\end{proof}

\section{Related~Work}
Data mining is used for a variety of purposes: from discovering relationships among attributes in big databases~\cite{srivastava2011performance}, to the classification of knowledge contained in heterogeneous data streams~\cite{farid2013mining}, to modeling customers' loyalty from purchasing behaviour~\cite{bunnak2015applying}, to newsworthy event anticipation from social medial posting patterns~\cite{jain2016real}, to fake profiles detection in social media~\cite{kadam2022social}.
While knowledge discovery is one of the main goals of data mining, 
the latter has been very scarcely used for the more specific goal of discovering new mathematics.
The only effort in this direction we are aware of is in~\cite{colton1999refactorable,colton2001mathematics}, where only the OEIS was mined. 
In our work, the crucial difference is the combined mining of both the OEIS and the huge Google and Google Scholar data sets. On one hand, this makes the potential set of interesting relationship between mathematical entities order of magnitudes bigger; on the other hand, relying only on textual comparison, our approach requires bigger human intervention to examine and prove the discovered potential relationships.

\section{Conclusions}
\label{RefSectConcl}
Cued by a match obtained by web-searching data mined from the OEIS, we showed that there is a one-to-one correspondence between event structures and full graphs: see Theorem~\ref{RefLmBij} and Corollary~\ref{RefLmCard}, derived from Theorem~\ref{RefLmRepr}.

The latter is an original addition to a family of fundamental theorems relating basic algebraic structures to elementary mathematical constructions by establishing that the two entities exhibit the same behaviour, and commonly referred to as representation theorems.
Among the best known instances are Birkhoff's representation theorem~\cite
{davey2002introduction} characterizing finite distributive lattices through set-theoretical union and intersection, Stone's and Birkhoff's theorems offering related representations for Boolean algebras~\cite
{givant2008introduction}, Cayley's theorem implementing groups as permutations~\cite
{mac1999algebra}.
Many fundamental abstract structures historically arose from abstracting the properties of some operations on more concrete objects (e.g., join and meet in distributive lattice incarnate union and intersection), which can therefore be regarded as prototypical examples for the relevant structures.
Typically, a representation theorem closes the circle and goes back from the abstract structure to the prototypical example, showing that it can be used to represent any instance of the abstract structure.
In the case of Theorem~\ref{RefLmRepr}, however, one certainly cannot say that $\supseteq$ and disjoint intersection are prototypical examples for the relations of causality and conflict of an ES, mainly because, historically, ES developed in the setting of concurrency theory, largely detached from set-theoretical notions.
It is probably this fortuity which prevented that theorem, 
and consequently results~\ref{RefLmBij} and~\ref{RefLmCard} linking ESs and full graphs, from being stated earlier.%
\footnote{There are representation theorems relating to \emph{categories} of ESs, at a more abstract level than the present work. 
E.g.~\cite{winskel1999event}.}
It is likely that similar unfavourable, historical circumstances prevent 
further 
discoveries
linking seemingly mutually unrelated mathematical theories: we believe that data mining and AI approaches are worth being further pursued in such cases, and this paper is a proof of concept supporting this claim.
%
Given the way our theorems were obtained, the point of making sure, and of convincing the community of their correctness is of particular importance.
For this reason, we produced a formal proof of our results, and successfully checked its correctness with the \I{}/HOL proof assistant.
A separate paper is being written to describe this formalisation effort, the corresponding challenges, ideas and solutions, and will be posed to the automated reasoning community to gauge the interest in a potentially fruitful, novel intersection between subdomains of AI. 

We conclude with some cues for future work.
One limitation needing attention is the human role in parsing the matches obtained in Section~\ref{RefSectMining}: while we believe that, to find interesting theorems, human intervention is key, there is space for improvement in pruning the irrelevant matches and better leveraging the huge amount of knowledge available through web searches.
For example, NLP techniques could improve the crude keyword-based approach 
of Section~\ref{RefSectMining}
to single out mathematical concepts.
Another, more technical, limitation in need to be mitigated is the difficulty of inserting mathematical manipulations in the web search process; this is related to the plain-text interface used in web search queries, and to the fact that we have no control on the transformations applied by the web searching platform over the set of indexed documents (which would probably be too big to transform even if there were some form of control).

More specifically to the original theorems introduced in this paper, one obvious direction of development is the extension of Theorem~\ref{RefLmRepr} to the infinite case, in a way analogous to how Priestley's representation theorem generalises Birkhoff's~\cite[Theorem~11.23]{davey2002introduction}.
Using this generalisation as a guidance, this will probably require non-trivial conceptual leaps (of a scale analogous to the interpretation of Stone's theorem Priestley devised to obtain her result).

\bibliographystyle{IEEEtran}
\bibliography{13}

\end{document}